\newcommand{\be}{\begin{equation}}
\newcommand{\ee}{\end{equation}}
\newcommand{\bea}{\begin{eqnarray}}
\newcommand{\eea}{\end{eqnarray}}
\newcommand{\ba}[1]{\begin{array}{#1}}
\newcommand{\ea}{\end{array}}
\documentclass[preprint,12pt]{elsarticle}

\usepackage{epsfig}
\usepackage{amsmath}
\usepackage{amssymb}
\usepackage{mathrsfs}
\usepackage{color}
\usepackage{rotating}

\journal{OPTICS COMMUNICATIONS}

\begin{document}

\begin{frontmatter}



\title{Photon-Photon Correlations With a V-Type Three-Level System Interacting with Two Quantized Field Modes}



\author{Arpita Pal}
\author{Bimalendu Deb\corref{cor1}}
\address{School of Physical Sciences, Indian Association for the Cultivation of Science, Jadavpur, Kolkata 700032, India}
\ead{msbd@iacs.res.in}
\cortext[cor1]{Corresponding author. Address: School of Physical Sciences, Indian Association for the Cultivation of Science, Jadavpur, Kolkata 700032, India, Tel: +91 33 2473 4971 (Ext 1201), Fax:+91-33-2473 2805  }
\def\zbf#1{{\bf {#1}}}
\def\bfm#1{\mbox{\boldmath $#1$}}
\def\hf{\frac{1}{2}}

\begin{abstract}
We carry out a model study on the interaction of a $V$-type three-level emitter with two quantized cavity modes which are weakly driven by two classical fields. The emitter may be an atom or a molecule with nondegenerate upper levels in general. The lifetimes of the two upper levels are assumed to be much longer than the lifetime of the cavity photons. We calculate the two-time second order coherence function, namely Hanbury Brown-Twiss function $g^{(2)}(\tau)$ where $\tau$ is the time delay between the two modes. We analyze the photon-photon correlations between the two cavity modes in terms of $g^{(2)}(\tau)$. The variation of $g^{(2)}(\tau)$ as a function of $\tau$ exhibits collapse and revival type oscillations as well as quantum beats for relatively short $\tau$ while in the limit $\tau \rightarrow \infty$, $g^{(2)}(\tau) \simeq 1$. We further show that the two cavity field modes are entangled when $g^{(2)}(0) > 2$. We develop a dressed state picture with single photon in each mode to explain the results and use the negativity of the partial transpose of the reduced field density matrix to show the entanglement between field modes. The model presented in this paper may be useful for generating entangled photon pairs, and also manipulating photon-photon correlations with a cavity QED set up.

\end{abstract}


\end{frontmatter}


\section{Introduction}
\label{intro}
One of the major goals  of current research in quantum optics is the generation of non-classical states of light i.e. to make efficient sources of single photons as well as correlated or entangled pairs of photons for a variety of applications in the emerging area of quantum technology \cite{photon_quantum_tech}. Ability to manipulate quantum correlations and to generate an interaction between a pair of photons is particularly important for photonic quantum information processing and communication \cite{kimble2008}. In this context, cavity quantum electrodynamics (CQED) \cite{berman:book, doherty:science2002}, where atoms interact with a quantized electromagnetic field inside a cavity, constitute one of the most useful systems to engineer entanglement \cite{harochermp2001, agarwalprl2003, harochebook}, photon blockade \cite{imamoglu:prl1997, kimble:nature2005, agarwalpra2017} and a variety of quantum correlations \cite{orozcoprl1998, tanpra2004,rempe:nature2008,rempe:prl2008, kochprl2011, murrnature2011, 
lukinnp2014}. Apart from the extensive use of two-level systems in CQED \cite{kimble:nature2005, rempe:prl2008, kochprl2011} or quantum dots in semiconductor\cite{q_dot}, nitrogen vacancy in diamond \cite{nv} to generate non-classical lights, $\varLambda$-type three-level atoms or multilevel $N$ - systems with two lower levels being ground-state sub-levels have been widely employed for exploring a class of coherent optical effects based on the long-lived atomic coherence. At the heart of this atomic coherence lies the dark state resonance \cite{arimondo, dark_resonance, eit:rmp, eit:cohen} that effectively decouples the lossy excited third level from the coherent superposition of the two lower levels, leading to vanishing light absorption. This is the essence of electromagnetically induced transparency (EIT) \cite{eit:rmp}. $\varLambda$-type three-level atoms in cavities have been used to study optical nonlinearity \cite{rempeprl2000, lukinprl2000,xiaoprl2001,rempeprl2002}. On the other hand, $V$ - type 
three-level systems have not found much applications in coherent spectroscopy, primarily because of the existence of two lossy excited states unlike that in $\varLambda$ system. Nevertheless, three-level $V$ - system in a cavity or a waveguide has, of late, attracted a considerable research interest for inducing giant Kerr nonlinearities \cite{ficekjpb2009}, cavity-enhanced single-photon emission \cite{schumacherprb2018}, generation of entanglement \cite{ajiki:pss2009, safaripramana2013}, chiral optics \cite{chiral} and many other related phenomena. 

Here we study photon-photon correlations with a $V$ - type three-level system inside a two-mode cavity. We characterize the correlation in terms of two-time Hanbury Brown-Twiss (HBT) \cite{hanbury} correlation function $g^{(2)}(\tau)$, where $\tau$ stands for the time delay between the two field modes. The correlation between any two light fields can be quantified by the $g^{(2)}(\tau)$ function \cite{scully:pra2005}. At the outset, we consider a general $V$-type emitter which may be an atomic or molecular system with two degenerate or non-degenerate excited states. The cavity can support two modes that are near resonant to the two optical dipole transitions. We assume that the excited levels have much longer lifetime than the cavity-field lifetime. In our model, the coherence time is mainly limited by the inverse of the cavity damping constant since the atomic damping is assumed to be negligible during cavity relaxation time. We derive dressed-state picture of our model. We then write the master equation in 
terms of joint atom-field basis and solve the equation in steady state. We calculate $g^{(2)}(\tau)$ for the steady-state density matrix making use of the quantum regression theorem \cite{quantumregression}. The anharmonicity of the eigen energies of the dressed $V$ - system can influence photonic correlations in different parameter regimes.

Our results show that by tuning the detuning of one of the drive fields keeping other parameters fixed, one can switch over the photon-photon correlation from  bunching to antibunching  or vice versa. We demonstrate here that the model emitter-cavity system can behave as a generator of entangled photon-pairs $g^{(2)}(0)> 2$ under proper detuning of the probe lasers. On the other hand, the same system can be used as  single-photon source for different probe laser frequencies in strong coupling regime. We explain our results from the point of view of two-mode cavity-dressed picture of $V$ - system where the dressed levels are being probed by the two weak classical drives. In this context, it is worth mentioning that in recent times, many groups have theoretically demonstrated the super-thermal photon bunching in a quantum-dot-based two-mode microcavity laser \cite{leymann:pss2013}, enhanced two-photon emission from a dressed biexciton \cite{Valle:njp2015}, optical switching of cross intensity correlation in 
cavity EIT \cite{Rao:jpb2017}, quantum cavity-assisted spontaneous emission as a single-photon source using two cavity modes \cite{khanbekyan:pra2018} and many related articles in the context of correlation spectroscopy. 

The paper is organized in the following way. In section 2 we present our model consisting of a $V$-type three level system having long-lived or metastable excited states interacting with a double crossed cavity setup as shown in Fig.\ref{schematic}, where each cavity supports only one mode of oscillation frequency. We describe the system Hamiltonian and find emitter-field dressed energy eigenstates which form an anharmonic ladder (see Fig.\ref{ladder}). In section 3 we discuss about the construction of the density matrix and the solution of the master equation. In section 4 we calculate and describe the two-time correlation function $g^{(2)}(\tau)$. Finally, we discuss our numerical results in section 5 and then conclude in section 6.

\section{The Model}
\begin{figure}
\includegraphics[width=\linewidth]{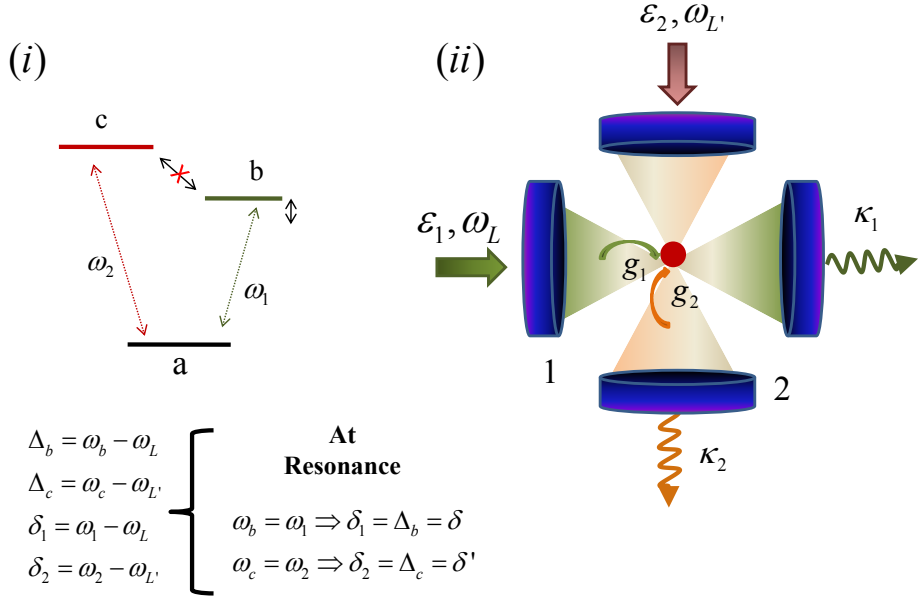}
\caption{(color online)(i) The level diagram of a $V$-type system, a is the ground state and b and c are two excited states having frequencies $\omega_b$ and $\omega_c$ respectively. $\omega_1$ and $\omega_2$ are cavity mode frequencies which are near resonant to the transition $a\leftrightarrow b$ and $a\leftrightarrow c$ respectively. (ii) A possible schematic setup consisting of a double crossed cavity setup  where the cavity waists overlap each other, each cavity (1 and 2) supports only one mode of oscillation ($\omega_1$ or $\omega_2$). Both field modes 1 and 2 are weakly driven by probe lasers having frequencies $\omega_{L}$ and $\omega_{L'}$ respectively. $\varepsilon_1$ and $\varepsilon_2$ are the corresponding driving rate of the two probe lasers. ${\rm g}_1$ and ${\rm g}_2$ are the atom-field coupling parameters and $\kappa_1$ and $\kappa_2$ are the respective cavity decay rates. Expressions for the detuning of the excited levels ($\Delta_b, \Delta_c$) and field modes ($\delta_1, \delta_2$) with respect to the probe fields are given. At atom-cavity resonances for both of the modes the detuning parameters are $\delta$ and $\delta'$.}
\label{schematic}
\end{figure}
We consider a V-type three-level quantum emitter which can be an atomic or a molecular system in general, coupled to a cavity set up as shown in Fig.\ref{schematic}. The cavity supports two modes of oscillation frequencies $\omega_1$, $\omega_2$ which are near resonant to the transitions $a\leftrightarrow b$ and $a\leftrightarrow c$. Both cavity modes 1 and 2 are weakly driven by some classical probe lasers having frequencies $\omega_{L}$ and $\omega_{L'}$ respectively. Let $\varepsilon_1$ and $\varepsilon_2$ be the corresponding driving rates of those two lasers. Here we assume that the excited levels are much more long-lived compared to the cavity modes. Our assumption is in contrast to the usual optical cavity QED situation where atomic damping is generally not negligible compared to  the cavity damping. The motivation for this assumption in our model stems from the recent experimental developments in high-precision  spectroscopy with cold atoms \cite{enomotoprl2008, kitagawapra2008} leading to the formation of excited atomic or molecular states with long lifetimes of the order of millisecond, while the typical lifetime of high-Q optical cavity is on the order of microsecond. Ultracold two-electron atoms  such as Yb \cite{enomotoprl2008, takahashipra2016} and Sr \cite{sr_longlived} can be employed to access  metastable or long-lived excited states via spin-forbidden inter-combination electric dipole transitions. Though such transitions are being considered for many applications including atomic clocks \cite{atomic:clock}, they are most probably not being considered so far for any interesting cavity QED (CQED) study. If the damping of the excited levels of $V$-type three-level system becomes negligible, then it is as good as the $\Lambda$-type three-level atoms, as far as dark-state resonance and associated coherent phenomenon such as EIT \cite{eit:rmp} is concerned. However, there is a significant difference between the dark resonances in the two systems: In case of $\Lambda$ - system the dark state arises due the destructive interference between the  two \emph{absorptive} transitions while in case of $V$-type system the dark state will be formed as a coherent superposition of the two excited states  due to the destructive interference between two \emph{emission} pathways. In this paper we do not directly address the problem of dark resonance in $V$-type system, but we focus our attention into the photon-photon correlations in the $V$-type system when both the fields that connect the two dipole transitions are quantized and the cavity modes are weakly driven by two classical fields.

\subsection{Hamiltonian}
The Hamiltonian can be written as
\bea
H = H_0 + H_{int}+ H_{drive}
\eea
with
\bea
H_0 &=& \hbar \omega_b \mid b\rangle \langle b \mid + \hbar \omega_c \mid c\rangle \langle c \mid + \hbar \omega_1 a^{\dag}_1 a_1 + \hbar \omega_2 a^{\dag}_2 a_2\nonumber\\
H_{drive} &=& \hbar \varepsilon_1 (a_1 e^{i\omega_Lt} + a^{\dag}_1 e^{-i\omega_{L}t}) + \hbar \varepsilon_2 (a_2 e^{i\omega_{L'}t} + a^{\dag}_2 e^{-i\omega_{L'}t}) 
\label{original_h}
\eea
where $\omega_b$ and $\omega_c$ are the eigen energies of the level b and c, respectively. $a_{1(2)}$ represents the annihilation operator of field mode 1 (2). $H_{int}$ represents interaction hamiltonian between the three level system and the field modes. Under electric dipole and rotating wave approximations (RWA), it is given by
\bea
H_{int} =\left( \hbar {\rm{g_1}} a_1 A^{+}_b +  \hbar {\rm{g_2}} a_2 A^{+}_c \right) + H.c. 
\eea
where ${\rm{ g_1}}$ and ${\rm{ g_2}}$ are the coupling constants.  $A^+_{b(c)}$ represents the atomic raising operator $\mid b(c) \rangle \langle a \mid$. In a ``rotating'' reference frame of a time-dependent unitary transformation
 \bea
 W(t)={\rm exp}\left[-it \left(\omega_L \mid b \rangle\langle b \mid + \omega_{L'} \mid c \rangle\langle c \mid + \omega_L a^{\dag}_1 a_1 + \omega_{L'} a^{\dag}_2 a_2\right)\right]
 \eea
 the original Hamiltonian of Eq.(\ref{original_h}) is reduced to
\bea
{\cal H} = {\cal H}_0 + {\cal H}_{int}+ {\cal H}_{drive}
\eea
with
\bea
{\cal H}_0 &=&  \hbar \Delta_b \mid b\rangle \langle b \mid + \hbar \Delta_c \mid c\rangle \langle c \mid + \hbar \delta_1 a^{\dag}_1 a_1 + \hbar \delta_2 a^{\dag}_2 a_2\nonumber\\
{\cal H}_{int} &=&  \hbar \left({\rm{g_1}} a_1 A^{+}_b  + {\rm{g_1}} a^{\dag}_1 A^{-}_b \right) +  \hbar \left( {\rm{g_2}} a_2 A^{+}_c  + {\rm{g_2}} a^{\dag}_2 A^{-}_c\right)\nonumber\\
{\cal H}_{drive} &=& \hbar \varepsilon_1 (a_1 + a^{\dag}_1) + \hbar \varepsilon_2 (a_2 + a^{\dag}_2)
\label{eq1}
\eea
where $\Delta_b = \omega_b - \omega_L$, $\Delta_c = \omega_c - \omega_{L'}$, $\delta_1 = \omega_1 - \omega_L$ and $\delta_2 = \omega_2 - \omega_{L'}$. We consider for our study the cavity mode fields are at resonance with the two dipole allowed transitions, i.e.
\bea
\Delta_b &=& \delta_1= \delta \nonumber \\
\Delta_c &=& \delta_2 = \delta'
\label{2preso}
\eea
Hence the detuning parameters $\delta$ and $\delta'$ are only guided by the frequencies of the classical drive lasers.
\begin{figure}
\includegraphics[width=\linewidth]{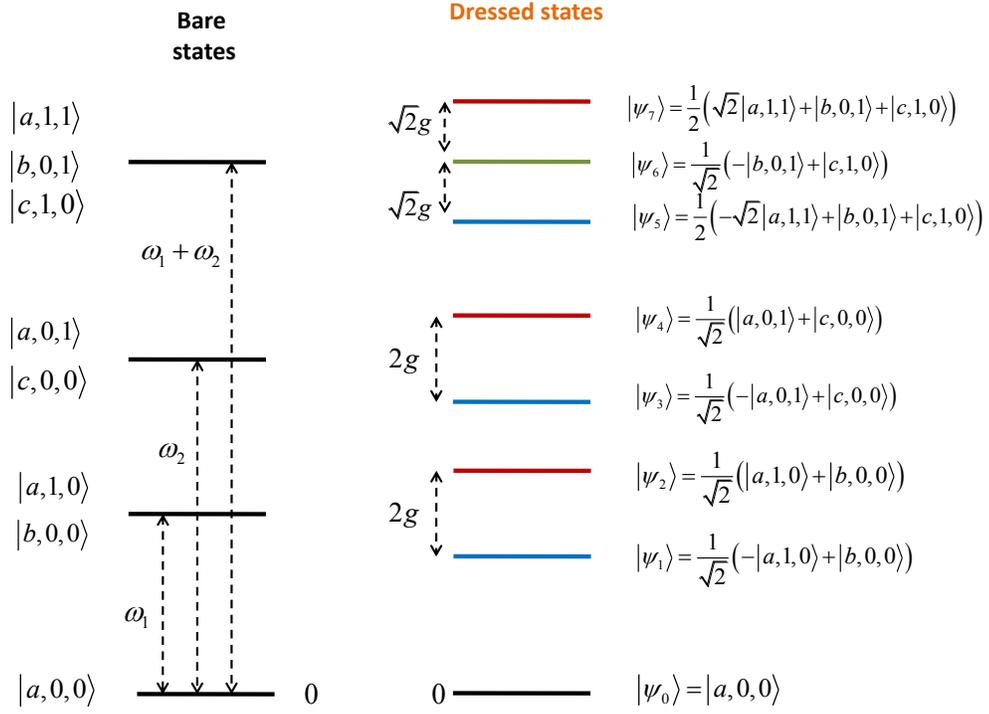}
\caption{(color online) Energy levels for the system upto $n_1 = n_2 = n = 1$ when $\delta = \delta' = 0$ and ${\rm g}_1 = {\rm g}_2$. The bare states in the system-mode photon resonance are shown in the left side  and in the right side the eigen states with corresponding energy splitting in rotating frame are shown for photon sectors (0,0), (1,0), (0,1), (1,1), where upto one photon exists in each mode.}
\label{ladder}
\end{figure}
\subsection{Dressed states}
Let us suppose that the modes 1 and 2 are initially in photon-number states with $n_1$ and $n_2$ photons, respectively. Then under RWA and in the absence of any dissipation, we can restrict the quantum dynamics within three atom-field joint bare basis which are
\bea
\mid a, n_1, n_2 \rangle,\hskip 1cm \mid b, n_1-1, n_2 \rangle,\hskip 1cm \mid c, n_1, n_2-1 \rangle
\eea
We call this ($n_1, n_2$) photon sector. For simplified analytical solution we consider both modes have equal number of photons $n_1 = n_2 = n$. For very weak driving and appropriate detuning of the two drive fields, we can restrict ourselves in two-photon subspace with each mode containing one photon only, (i.e. upto n = 1). The detunings should be much smaller than the frequency gap between the highest dressed level (1,1) photon sector and (1,2) or (2,1) photon sector so that none of the driving fields can effectively excite any higher dressed level. These conditions are the two-mode $V$-system counterpart of those required to realize an effective two-state system of Jaynes-Cummings model as first introduced by Tian and Carmichael \cite{carmichaelpra1992} who call it as ``dressing of the dressed states''. Hence the first set of bare states will be
\bea
\mid a, 1, 1 \rangle,\hskip 1cm \mid b, 0, 1 \rangle,\hskip 1cm \mid c, 1, 0 \rangle.
\eea
When the system is in ground state, then one photon is present in each of the modes. The eigen energies of the Hamiltonian (\ref{eq1}) satisfying the condition of  Eq.(\ref{2preso}) without ${\cal H}_{drive}$ are
\bea
E_5 &=& \hbar (\delta + \delta')-\sqrt{({\rm g}^2_1+{\rm g}^2_2)}\nonumber\\
E_6 &=& \hbar (\delta + \delta')\nonumber\\
E_7 &=& \hbar (\delta + \delta')+\sqrt{({\rm g}^2_1+{\rm g}^2_2)}
\eea
and the corresponding eigenvectors are
\bea
\mid \psi_5\rangle &=& \frac{{\rm g}_2}{\sqrt{2({\rm g}^2_1+{\rm g}^2_2)}} \left(-\frac{\sqrt{{\rm g}^2_1+{\rm g}^2_2}}{{\rm g}_2}\mid a \rangle + \frac{{\rm g}_1}{{\rm g}_2}\mid b \rangle + \mid c \rangle \right)\nonumber\\
\mid \psi_6\rangle &=& \frac{{\rm g}_1}{\sqrt{{\rm g}^2_1+{\rm g}^2_2}} \left(-\frac{{\rm g}_2}{{\rm g}_1}\mid b \rangle + \mid c \rangle \right)\nonumber\\
\mid \psi_7\rangle &=& \frac{{\rm g}_2}{\sqrt{2({\rm g}^2_1+{\rm g}^2_2)}} \left(\frac{\sqrt{{\rm g}^2_1+{\rm g}^2_2}}{{\rm g}_2}\mid a \rangle + \frac{{\rm g}_1}{{\rm g}_2}\mid b \rangle + \mid c \rangle \right)
\eea
when ${\rm g}_1 = {\rm g}_2 = {\rm g}$, i.e. both modes have same coupling the corresponding eigenvectors are
\bea
\mid \psi_5\rangle &=& \frac{1}{2} \left(-\sqrt{2}\mid a,1,1 \rangle + \mid b,0,1 \rangle + \mid c,1,0 \rangle \right)\nonumber\\
\mid \psi_6\rangle &=& \frac{1}{\sqrt{2}} \left(-\mid b,0,1 \rangle + \mid c,1,0 \rangle \right)\nonumber\\
\mid \psi_7\rangle &=& \frac{1}{2} \left(\sqrt{2}\mid a,1,1 \rangle + \mid b,0,1 \rangle + \mid c,1,0 \rangle \right)
\eea
When there is at most one photon in either of the modes the atom-field interaction causes the degenerate pair of bare states to split and form dressed states as in Jaynes-Cumming Hamiltonian for two-level system. The lowest 4 dressed energies are
\bea
E_1 &=& \hbar (\delta - {\rm g})\nonumber\\
E_2 &=& \hbar (\delta + {\rm g})\nonumber\\
E_3 &=& \hbar (\delta' - {\rm g})\nonumber\\
E_4 &=& \hbar (\delta' + {\rm g})
\eea
and the corresponding eigenvectors are
\bea
\mid \psi_1\rangle &=& \frac{1}{\sqrt{2}} \left(-\mid a, 1, 0 \rangle + \mid b, 0, 0 \rangle\right) \nonumber\\
\mid \psi_2\rangle &=& \frac{1}{\sqrt{2}} \left(\mid a, 1, 0 \rangle + \mid b, 0, 0 \rangle\right)\nonumber\\
\mid \psi_3\rangle &=& \frac{1}{\sqrt{2}} \left(-\mid a, 0, 1 \rangle + \mid c, 0, 0 \rangle\right) \nonumber\\
\mid \psi_4\rangle &=& \frac{1}{\sqrt{2}} \left(\mid a, 0, 1 \rangle + \mid c, 0, 0 \rangle\right)
\eea
It is to be noted that the eigenvectors become detuning-independent as we work in atom-cavity resonance condition (Eq.(\ref{2preso})). The total bare and dressed state level diagram for photon number space upto $n =1$ is shown in Fig.\ref{ladder}. If we consider $n_1 = 1$ and $n_2 = 2$, then the set of bare states will be
\bea
\mid a, 1, 2 \rangle,\hskip 1cm \mid b, 0, 2 \rangle,\hskip 1cm \mid c, 1, 1 \rangle
\eea
The dressed eigen energies are
\bea
E_8 &=& \hbar (\delta + 2\delta')-\sqrt{3}{\rm g}\nonumber\\
E_9 &=& \hbar (\delta + 2\delta')\nonumber\\
E_{10} &=& \hbar (\delta + 2\delta')+\sqrt{3}{\rm g}
\eea
Hence if $\mid\mid \delta'\mid - \sqrt{3}{\rm g}\mid >> \kappa$, it justifies the restriction of the photon numbers to $(1,1)$ fock space.

\section{Density matrix and its solution}
To look into the system dynamics in more detail we opt over the density matrix equation in Linblad form
\bea
\frac{d\rho_{sf}}{dt} = {\cal L}\rho_{sf} = -\frac{i}{\hbar}\left[{\cal H}_{sf},\rho_{sf}\right] +\sum_{i=1,2} \frac{\kappa_i}{2}\left(2 a_i\rho_{sf}a^{\dag}_i - \rho_{sf} a^{\dag}_i a_i - a^{\dag}_i a_i \rho_{sf} \right)
\label{eq3}
\eea
$\rho_{sf}$ the the density matrix of joint atom-field system, $\cal L$ is called as the Liouville super operator. $\kappa_i$ denotes the decay rate of field mode i (i = 1, 2). As mentioned earlier, here we consider that the excited levels of $V$-type system are much more long lived compared to the cavity lifetime. For sake of simplicity of numerical analysis, we consider both atom-field coupling strengths are equal, i.e. $\rm{g_1} = \rm{g_2} = \rm{g}$, driving rates are also considered to be same i.e. $\varepsilon_1 = \varepsilon_2 = \varepsilon$ and the two cavity modes experience same decay rate $\kappa_1 = \kappa_2 = \kappa$. The bare states for our system as mentioned earlier are
\bea
\mid 0\rangle &=& \mid a, 0, 0\rangle, \mid 1\rangle = \mid a, 1, 0\rangle, \mid 2\rangle = \mid b, 0, 0\rangle, \mid 3\rangle = \mid a, 0, 1\rangle,\nonumber\\
 \mid 4\rangle &=& \mid c, 0, 0\rangle, \mid 5\rangle = \mid a, 1, 1\rangle, \mid 6\rangle = \mid b, 0, 1\rangle, \mid 7\rangle = \mid c, 1, 0\rangle
\label{bare}
\eea
We form the density matrix with these eight bare basis states. With normalization condition, the Eq.(\ref{eq3}) leads to 63 coupled first order differential equations. We cast theses coupled equations into a matrix form which is useful for applying quantum regression theorem \cite{quantumregression}. We then obtain the steady-state solutions of the density matrix and second order cross correlation function $g^{(2)}(\tau)$ around steady-state.
 
\section{Two-time photon-photon correlations}
The two-time second-order correlation function \cite{chu:book, jurczak:oc1995} can be written explicitly as
\bea
g^{(2)}(\tau) &=& \frac{\langle a^{\dag}_1(t) a^{\dag}_2(t+\tau)a_2(t+\tau)a_1(t)\rangle}{\langle  a^{\dag}_1(t) a_1 (t)\rangle \langle a^{\dag}_2(t+\tau) a_2 (t+\tau)\rangle}
\label{twotime}
\eea
Using quantum regression theorem \cite{quantumregression} we can calculate $g^{(2)}(\tau)$. The signatures of different types of photon correlations, such as bunching, antibunching can be obtained from $g^{(2)}(\tau)$. When $\tau = 0$, it reduces to same time second order cross correlation function $g^{(2)}(0)$ \cite{wallsmilburn} and is related to the photon statistics and Mandel's Q-parameter. At steady state it can be written as
\bea
g^{(2)}(\tau) &=& \frac{\langle a^{\dag}_1(t) a^{\dag}_2(t)a_2(t)a_1(t)\rangle}{\langle  a^{\dag}_1(t) a_1 (t)\rangle \langle a^{\dag}_2(t) a_2 (t)\rangle}\nonumber\\
&=& \frac{{\rm Tr} \left[a^{\dag}_1 a^{\dag}_2 a_2 a_1 \rho_{sf}(t\rightarrow \infty)\right]}{{\rm Tr} \left[  a^{\dag}_1 a_1\rho_{sf}(t\rightarrow \infty)\right] {\rm Tr} \left[  a^{\dag}_2 a_2\rho_{sf}(t\rightarrow \infty)\right]}
\label{g20}
\eea
where $\langle a^{\dag}_i a_i\rangle$ is the mean intra-cavity photon number of the $i^{th}$ cavity mode and $\rho_{sf}$ is the atom-field density matrix. For $\tau \sim 0$, the system response induces into photonic correlation ($g^2(0) > 1$) or anticorrelation ($g^2(0) < 1$) between the light fields. When $g^2(0) \gg 1$, there is highly correlated generation of photons in bunched form. i.e. bunching occurs. In contrast, when $g^2(0) < 1$, there occurs photon antibunching and as $g^2(0)\rightarrow 0$, the perfect photon blockade takes place, i.e. the system blocks the absorption of the second photon with large probability. The term ``photon blockade'' was first coined by A. Imamoglu and others in 1997 \cite{imamoglu:prl1997} in close analogy with the phenomena Coulomb blockade in solid-state system \cite{coulombblockade}.

Particularly for a weakly driven system, a photon-pair emission can also be described  by the differential correlation function $C^{(2)}(\tau)$ \cite{rempe:prl2008}, which at $\tau = 0$ is related to $g^{(2)}(0)$ by
\bea
C^{(2)}(0) = \langle a^{\dag2} a^2 \rangle - \langle a^{\dag}a \rangle^2 = \left[ g^{(2)}(0) - 1 \right] \langle a^{\dag}a \rangle^2
\label{c20}
\eea
The advantage of $C^{(2)}(\tau)$ is the reduced sensitivity to single photon excitations compared to $g^{(2)}(\tau)$. Hence it provides a more transparent measure of the probability to create two photons at once in cavity. For a coherent intracavity field $C^{(2)} (0) = 0$. And for weakly driven system $C^{(2)}(0)>0$ only when a two-photon state becomes significant compared to that of a single photon state. For antibunched light $C^{(2)}(0)<0$, i.e. negative.
\begin{figure}
\center
\includegraphics[width=\linewidth]{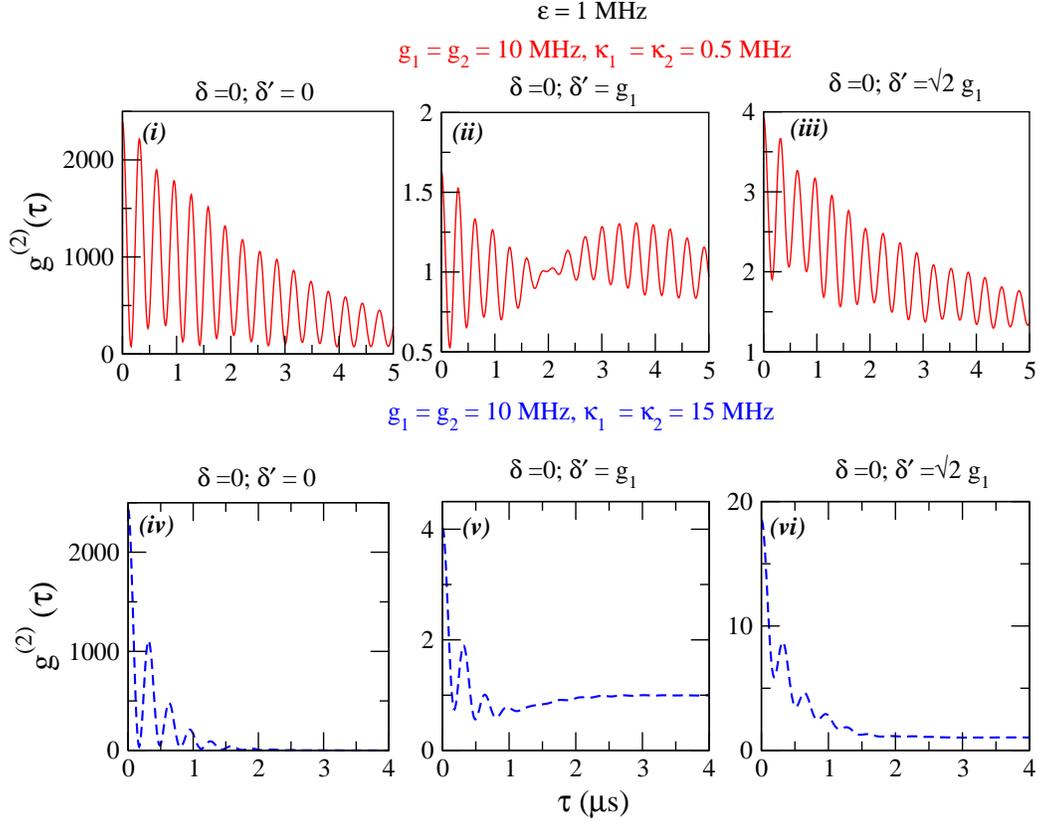}
\caption{(color online) The variation of two-time cross correlation function $g^{(2)}(\tau)$ as a function of time $\tau$ (in $\mu$s) is plotted for strong (red solid curve)(top three curves (i), (ii) and (iii)) and weak coupling (blue dashed curve) regime (bottom three curves (iv), (v) and (vi)). The strong and weak coupling model system parameters are given in the header and $\varepsilon = \varepsilon_1 = \varepsilon_2$.}
\label{g2tau}
\end{figure}
\begin{figure}
\center
\includegraphics[width=\linewidth]{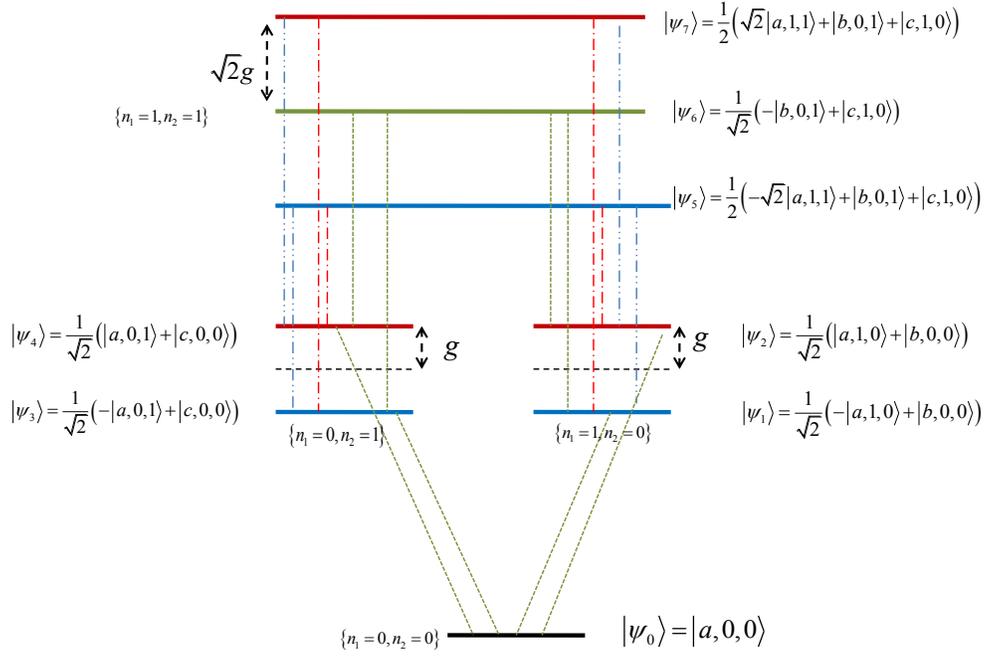}
\caption{(color online) Dressed states emission and different decay pathways of the mode photons from the dressed states are shown here. We consider that both modes experience same coupling ${\rm g}_1 = {\rm g}_2={\rm g}$. Green dashed lines are of photon frequency $\hbar [\delta(\delta')\pm {\rm g}]$, red dash-dotted lines are of frequency $\hbar [\delta(\delta')\pm(\sqrt{2}+1) {\rm g}]$ and blue dash-dot-dotted line represent to decay frequency $\hbar [\delta(\delta')\pm (\sqrt{2}-1){\rm g}]$.  }
\label{decay}
\end{figure}
\begin{figure}
\center
\includegraphics[width=\linewidth]{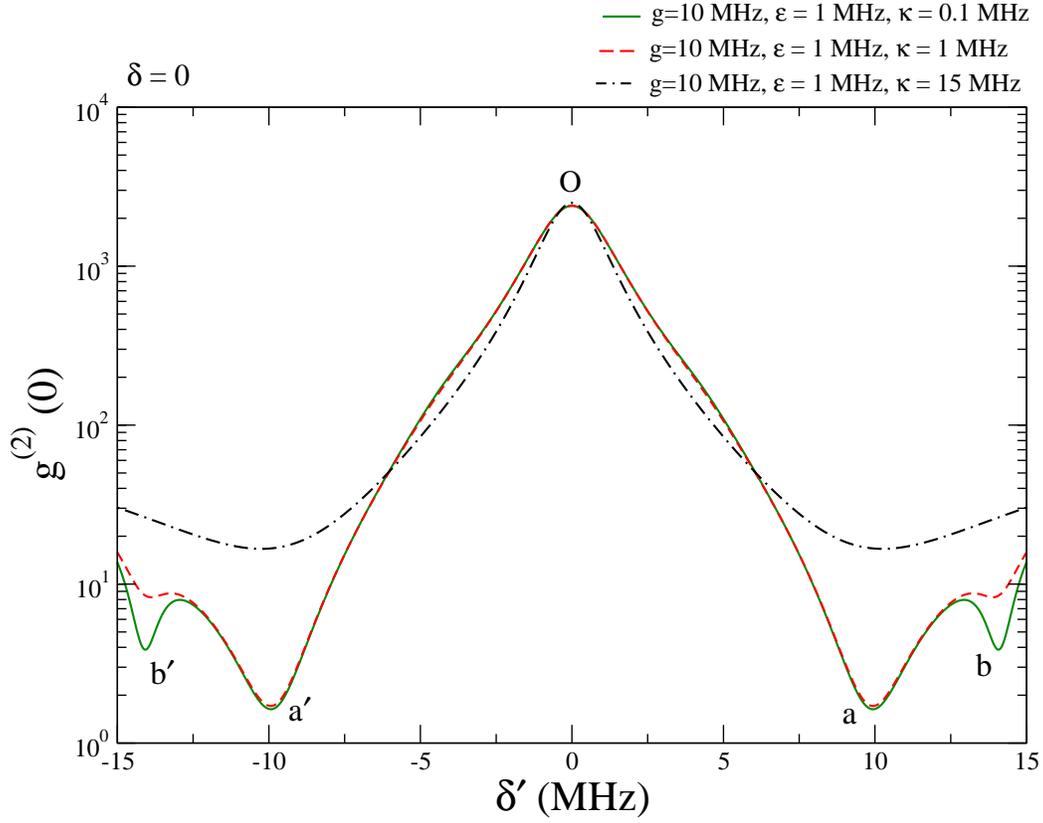}
\caption{(color online) Cross-correlation $g^{(2)}(0)$ is plotted as a fuction of $\delta'$ for three different values of $\kappa$ = 0.5 MHz (green solid), 1 MHz (red dashed), 15 MHz (dash-dot black) to access different coupling regime (strong to weak) keeping the other system parameters (${\rm g}$ = ${\rm g}_1$ = ${\rm g}_2$ = 10 MHz, $\delta = 0$ MHz and  $\varepsilon_1$ = $\varepsilon_2$ = $\varepsilon$ = 1 MHz) fixed.}
\label{g20_detuning}
\end{figure}
\begin{figure}
\center
\includegraphics[width=\linewidth]{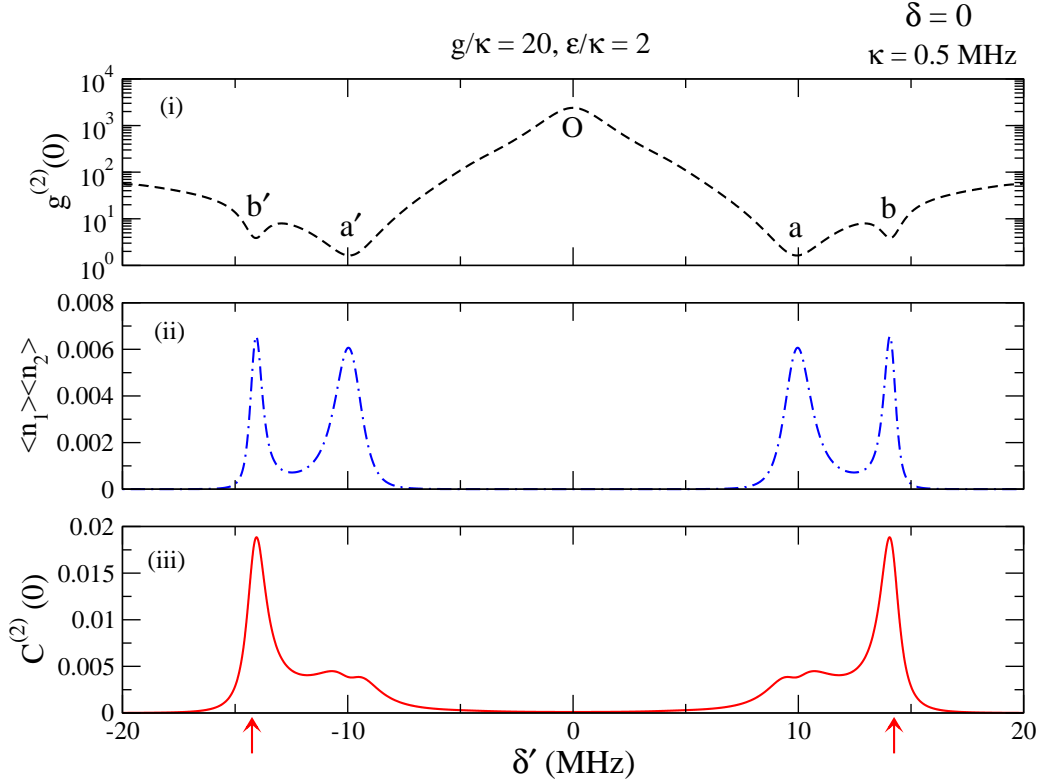}
\caption{(color online) $g^{(2)}(0)$, product of mean photon numbers of two modes ($\langle n_1 \rangle\langle n_2\rangle$) and $C^{(2)}(0)$ are plotted as a function of  one probe detuning $\delta'$ in (i), (ii) and (iii), respectively, when the other drive laser is on resonance at strong coupling regime. Two prominent peaks are observed in the $C^{(2)}(0)$ plot at the arrow-marked position of detuning $\delta =\pm \sqrt{2}{\rm g}$ (system being probed at dressed level $\mid \psi_5\rangle$ and $\mid \psi_7\rangle$), showing the presence of two-photon induced tunneling \cite{pintunnel} there. At $\delta'=0$, $\langle n_1 \rangle\langle n_2\rangle \sim 4.5 \times 10^{-8}$, $C^{(2)}(0) \sim 1.08 \times 10^{-4}$ and therefore $g^{(2)}(0)\approx \frac{C^{(2)}(0)}{\langle n_1 \rangle\langle n_2\rangle} \sim 10^3$.}
\label{2pustrong}
 \end{figure}
\section{Results and Discussions}
We have mentioned earlier that throughout our analysis we'll consider the emitter-cavity resonance condition, i.e. the cavity mode frequencies are on resonance with the two dipole-transition pathways. The detuning parameters $\delta'$ and $\delta$ will be solely guided by the external drive laser frequency. At first we show the variation in two-time cross correlation function $g^{(2)}(\tau)$ for different detuning of one probe laser $\delta' = \omega_c - \omega_{L'} = \omega_2-\omega_{L'}$ keeping the other probe laser on resonance, i.e. $\delta = \omega_b - \omega_L = \omega_1-\omega_L = 0$. In Fig.\ref{g2tau}, we plot the variation of $g^{(2)}(\tau)$ for both strong and weak coupling regimes considering weak driving $\varepsilon_1 = \varepsilon_2 = 1$ MHz and ${\rm g}_1 = {\rm g}_2 = {\rm g} = 10$ MHz. At two-photon resonance condition $\delta = \delta' = 0$ (Fig.\ref{g2tau}(i),(iv)), when $\tau = 0$, we observe the $g^{(2)}(\tau) \sim 2.4\times 10^3$ for both strong (red solid) and weak coupling (blue dashed) regime, which indicates the highly correlated emission of the two mode photons. At larger $\tau$ values, $g^{(2)}(\tau)$ shows modulated decay combined with different oscillations and finally $g^{(2)}(\tau) \rightarrow 1$ large time. Hence at large $\tau$ two photons will become uncorrelated. Next we plot $g^{(2)}(\tau)$ for $\delta = 0, \delta'={\rm g}_1 = {\rm g}_2$ for strong and weak coupling regimes in Fig.\ref{g2tau}(ii) and Fig.\ref{g2tau}(v) respectively. At strong coupling regime we observe that $g^{(2)}(\tau)$ oscillates from $<1$ to $>1$ as shown in Fig.\ref{g2tau}(ii), hence the photon correlation changes from super to sub-Poissonian statistics with $\tau$. We also observe collapse- and revival-type oscillations in strong coupling regime. In contrast, weak coupling regime does not exhibit collapse and revival instead shows faster oscillatory decay as $\tau$ increases. Similarly in strong coupling regime with increase of $\tau$ the correlation or anti correlation will die down and $g^{(2)}(\tau)\sim 1$. Similarly when $\delta = 0, \delta'=\sqrt{2}{\rm g}_1 = \sqrt{2}{\rm g}_2$, then we observe that at $g^{(2)}(0)\sim 4$, i.e. bunched emission and as $\tau $ increases it undergoes modulated decay with some oscillations and finally $g^{(2)}(\tau\rightarrow \infty) = 1.3 $, shows bunching only at strong coupling regime (Fig.\ref{g2tau}(iii)), where at weak coupling regime it goes to 1 at large $\tau$ (Fig.\ref{g2tau}(vi)), i.e. uncorrelated at long time. The decay profile of the $g^{(2)}(\tau)$ shows different oscillation frequencies at strong coupling regime. To analyze this oscillatory behavior  we consider different photonic decay pathways at the dressed state picture in Fig.\ref{decay}.  We find that the fast oscillation in each $g^{(2)}(\tau)$ plot (Fig.\ref{g2tau}) has frequency about 3 MHz. There is also other slow oscillations resembling quantum beats only observed in strong coupling regime. These quantum beats may be attributed to the interference between different transition pathways.

Next we plot $g^{(2)}(0)$ for different coupling regimes in Fig.\ref{g20_detuning}. At point ``O'' $\delta = \delta' = 0$ i.e. at two-photon resonance there is a high $g^{(2)}(0)$ value ($>> 2$), for all coupling regimes. This extreme large bunching may appear due to quantum interference between the driving field and atomic polarization \cite{carmichael:oc1991, xiao:pra1999} and also due to really small normalization constant i.e. $\langle n_1 \rangle\langle n_2\rangle \sim 4.5 \times 10^{-8}$, i.e. very small mean number of photons in each mode. At point a and a$'$ in Fig.\ref{g20_detuning}, $\delta' = \pm {\rm g}$ and $\delta = 0$, the driving fields become resonant with the transitions $\mid \psi_1 \rangle, \mid \psi_2 \rangle, \mid \psi_3 \rangle$ and $\mid \psi_4 \rangle$ (see Fig.\ref{ladder}). As a result there will be a coherence inside the system and that results into suppression of the correlated emission of the two photons. $g^2(0)\sim 1.5$ here also, i.e. the photon correlation is similar to that of all classical lights. At point b and b$'$ in Fig.\ref{g20_detuning}, $\delta' = \pm \sqrt{2}{\rm g}$ and $\delta = 0$, here also the driving field become resonant with the transition to the dressed states $\mid \psi_5 \rangle, \mid \psi_7 \rangle$ with energy spacing ``$\pm\sqrt{2}{\rm g}$'' as mentioned earlier and we observe bunching $g^{(2)}(0)\sim 4$.

Though $g^{(2)}$(0) function shows the presence of strong photon-photon correlations,  another parameter $C^{(2)}(0)$ can be a more appropriate choice to observe the two photon absorption or emission in pairs, because the normalization factor is taken care of in $C^{(2)}(0)$ and it'll essentially peak around photon induced tunneling \cite{pintunnel}. At strong coupling regime we have plotted $g^{(2)}(0)$, $\langle n_1 \rangle\langle n_2\rangle$ and $C^{(2)}(0)$ in Fig.\ref{2pustrong}. We can observe that at $\delta'  = \pm \sqrt{2}{\rm g}$ there is a positive peak or the maxima of $C^{(2)}(0)$, which indicates that at detuning  of the dressed levels $\mid \psi_5\rangle$ and $\mid \psi_7\rangle$ , which contains bare state $\mid a, 1, 1\rangle$, which essentially stimulate the absorption or emission of two photons at once as discussed in \cite{rempe:prl2008}. At this point the value of $C^{(2)}\sim 0.02$, which is small positive value but the value of $g^{(2)}(0) \sim 4$ (Fig. 3(iii)), indicates that the correlation is not thermal and essentially bunching occurs. At this very point, the product of the mean photon numbers is $\langle n_1 \rangle\langle n_2\rangle \sim$ 0.006, compared to this $C^{(2)}\sim 0.02$ is one order higher. Hence the positive peak indicates the generation of two photons at once in the cavity with maximum probability under the given conditions. At $\delta'= 0$, $\langle n_1 \rangle\langle n_2\rangle \sim 4.5 \times 10^{-8}$, which results into obtaining a really small $C^{(2)}(0) \sim 1.08 \times 10^{-4}$. At $\delta' = 0$, $C^{(2)}(0)$ becomes minimum as it is the difference of two small probabilities.

\begin{figure}
\center
\includegraphics[width=\linewidth]{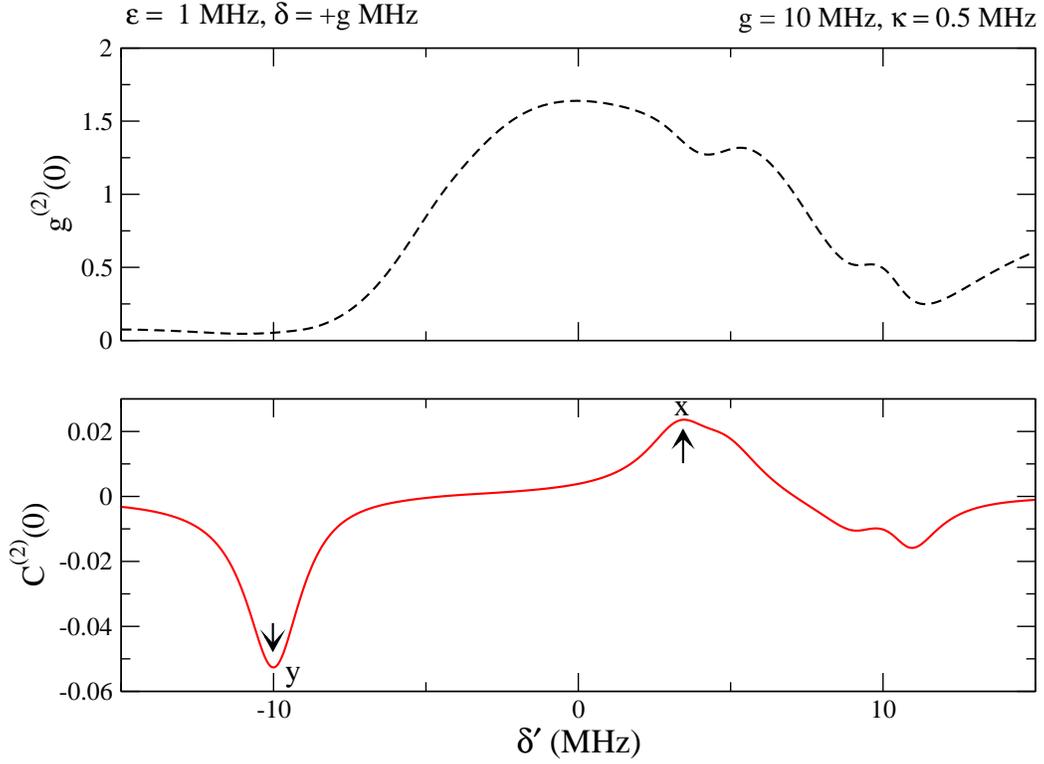}
\caption{(color online) $g^{(2)}(0)$ and $C^{(2)}(0)$ are plotted when one probe is kept fixed at $\delta = {\rm g}$ and the other parameters are mentioned at the top. At very strong coupling regime, the upper plot shows the $g^{(2)}(0)$ and the lower plot shows the differential correlation function $C^{(2)}(0)$ as a function of $\delta'$. At point ``x'', it shows bunching when the detuning is $\delta' = (\sqrt{2}-1){\rm g}$ and at point ``y'' when $\delta'= -{\rm g}$, it shows prominent deep $C^{(2)}(0)<0$, indicates strong antibunching.}
\label{blockade}
\end{figure}

After observing the correlated pair emission, we now want to see whether photon blockade is possible in our system for some range of parameters. Previously, photon blockade phenomena is mostly studied for identical mode photons, with two-level or multilevel systems \cite{imamoglu:prl1997,kimble:nature2005}. In our system, the two mode photons may be of different frequencies or polarization. We look for some range of parameters where we can limit the $g^{(2)}(0) \ll 1$ which implies photon blockade. We scan our system for a range of parameters and choose to probe it at $\delta = {\rm g}$ and vary $\delta'$. In Fig.\ref{blockade} the top plot shows the cross correlation function of the two-modes ($g^{(2)}(0)$) with $\delta'$ at very strong coupling regime, when the system is weakly driven. We observe there is a regime of antibunching where $g^{(2)}(0)\ll 1$. Now at the bottom plot in Fig.\ref{blockade} we plot $C^{(2)}(0)$ and observe that at $\delta' = -{\rm g}$ it shows prominent antibunching (point ``y'' in the plot). When the probe has $\delta = {\rm g}$ the dressed levels $\mid \psi_2 \rangle$ and $\mid \psi_4 \rangle$ are being probed. Then if a second photon arrives with $\delta' = \pm{\rm g}$, there is no excitation left, hence naturally the system will not absorb it until and unless it releases the first photon. This phenomena describes for obtaining such a low correlation function value $C^{(2)}(0)$ at $\delta' = \pm {\rm g}$. However in Fig.\ref{blockade} the antibunching at position ``y'' is more prominent. Another interesting observation we make that at point ``x'' bunching occurs and $\delta' = (\sqrt{2}-1){\rm g}$ here approximately at $\delta'\sim 4$ MHz here. When $\delta = {\rm g}$ the levels $\mid \psi_2 \rangle$ and $\mid \psi_4 \rangle$ are being probed in the dressed state picture. Now if a second photon arrives with $(\sqrt{2}-1){\rm g}$, then effective energy of the probe photons will be $\sqrt{2}{\rm g}$. With this energy the next excitation will take place to state $\mid \psi_7 \rangle$, which are separated $\sqrt{2}{\rm g}$ from the mid line (see Fig.\ref{ladder}). Hence  it is clear that two photons of energy $\rm g$ and $(\sqrt{2}-1){\rm g}$ is required for transition to state $\mid \psi_7 \rangle$. It is essentially a two-photon induced transition and hence two-photon will be emitted at once in the cavity and we find $C^{(2)}(0)>0$ and $g^{(2)}(0)>1$. If we set $\delta = -{\rm g}$, this photon induced tunneling will take place at about $\delta' = -(\sqrt{2}-1){\rm g}$ and then dressed state $\mid \psi_5 \rangle$ will be probed.

\begin{figure}
\center
 \includegraphics[width=\linewidth]{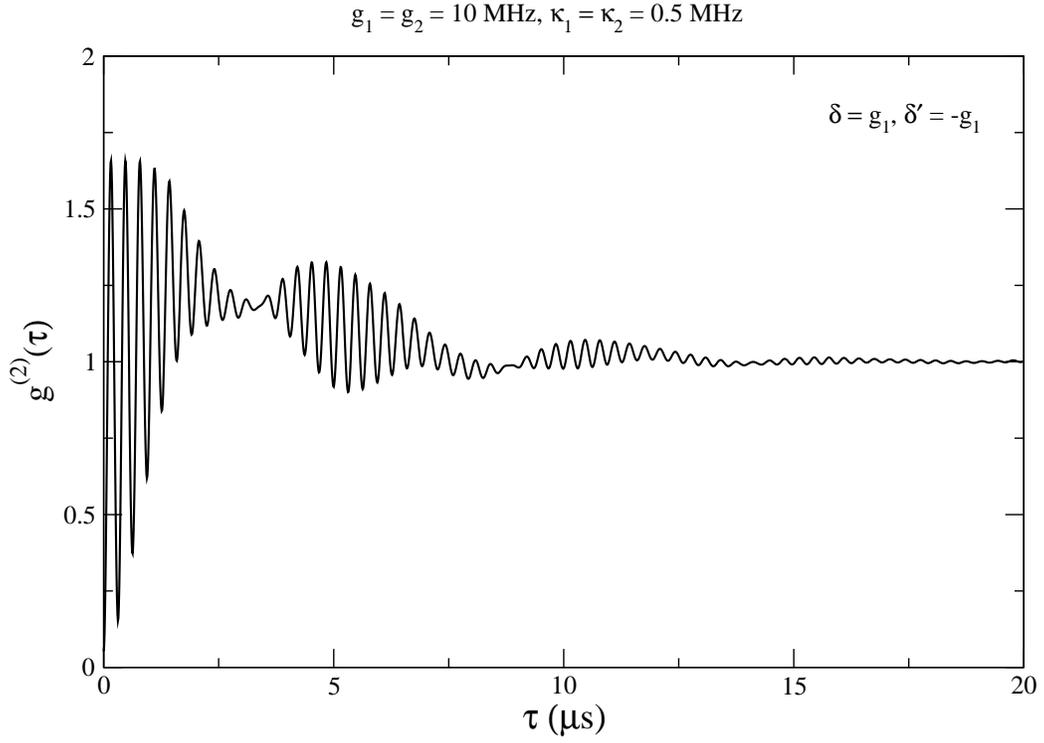}
\caption{ The two mode correlation function $g^{(2)}(\tau)$ is plotted against $\tau$ (in $\mu s$), when the cavity and drive lasers have detuning $\delta = {\rm g}_1 = {\rm g}_2$ and $\delta' = -{\rm g}_1 = -{\rm g}_2$ and the system is in strong coupling regime (parameters are mentions in the plot) under weak driving limit $\varepsilon_1$ = $\varepsilon_2$ = $\varepsilon = 1$ MHz.}
\label{g2tau_delg}
 \end{figure}

Next we analyze the antibunching of the photons at strong coupling regime ( at position ``y''  of Fig.\ref{blockade} with $\delta = $ and $\delta' = -{\rm g}$) with time delay $\tau$. We plot the variation of $g^{(2)}(\tau)$ with $\tau$ in Fig.\ref{g2tau_delg}. Here we can see at time $\tau$ = 0 $g^{(2)}(0)<1$, i.e mode photons are anti-correlated. With the increase of the time, correlation function starts oscillating, i.e the mode photons can be correlated or anti correlated at different time delays. As we can see in Fig.\ref{g2tau_delg} they exhibit collapse and revival type phenomena in the correlation. Eventually at long time limit the oscillations die down and $g^{(2)}(\tau\rightarrow\infty)\sim 1$.

To measure the nonclassicality of the photon states we study the violation of Cauchy-Schwarz inequality (CSI). At classical level the CSI for intensity correlation function has the form
\bea
\langle I(t) I(t+\tau)\rangle \leq \sqrt{\langle I^2(t)\rangle \langle I^2(t+\tau)\rangle}
\eea
In two mode quantum problem CSI can be formulated in terms of second-order correlation function $G^{(2)}_{ij} = \langle a^{\dag}_i a^{\dag}_j a_j a_i \rangle$ \cite{walls:pra1986, wallsmilburn, westbrook:prl2012, agarwal:qo}
\bea
G^{(2)}_{12} (\tau) \leq \left[G^{(2)}_{11} G^{(2)}_{22}\right]^{\frac{1}{2}}
\label{csi}
\eea 
Now for our model system the Fock states contain at most one photon for each mode  (see Eq.(\ref{bare})).  So for our model system, the two-mode Fock basis are $\mid 0,0 \rangle$, $\mid 1,0 \rangle$, $\mid 0,1 \rangle$ and $\mid 1,1 \rangle$. Using theses basis functions, one can obtain 
\bea
G^{(2)}_{11} &=& \langle a^{\dag 2}_1 a^2_1\rangle = Tr_f \left[\rho_f a^{\dag 2}_1 a^2_1 \right] = 0\\
G^{(2)}_{22} &=& 0 \nonumber
\eea
where $f$ stand for field states and $\rho_f$ is the reduced field density matrix. Hence the CSI in Eq.(\ref{csi}) for our system implies
 \bea
G^{(2)}_{12} (\tau) \leq 0
 \eea
Now we can write Eq.(\ref{csi}) in terms of two-time photon correlation function $g^{(2)}_{ij} (\tau)$ 
\bea
g^{(2)}_{12} (\tau) \leq \left[g^{(2)}_{11} g^{(2)}_{22}\right]^{\frac{1}{2}}
\eea
Hence for our system this inequality reduces to
\bea
g^{(2)}_{12} (\tau) \leq 0
\label{csi_1}
\eea 
For all our numerical results as described above $g^{(2)}(0) > 0$ , implying the violation of CSI. The condition in Eq.(\ref{csi_1}) shows that when $g^{(2)}_{12} (\tau)= g^{(2)}(0) = 0$ CSI will hold good and it will indicate classicality, but on the contrary $g^{(2)} (0) = 0$ implies perfect photon blockade condition which is essentially a non classical signature of the light field. Hence CSI may not be a good criterion to judge nonclassicality of our model system with such a low number of photons.
\begin{figure}
\center
 \includegraphics[width=\linewidth]{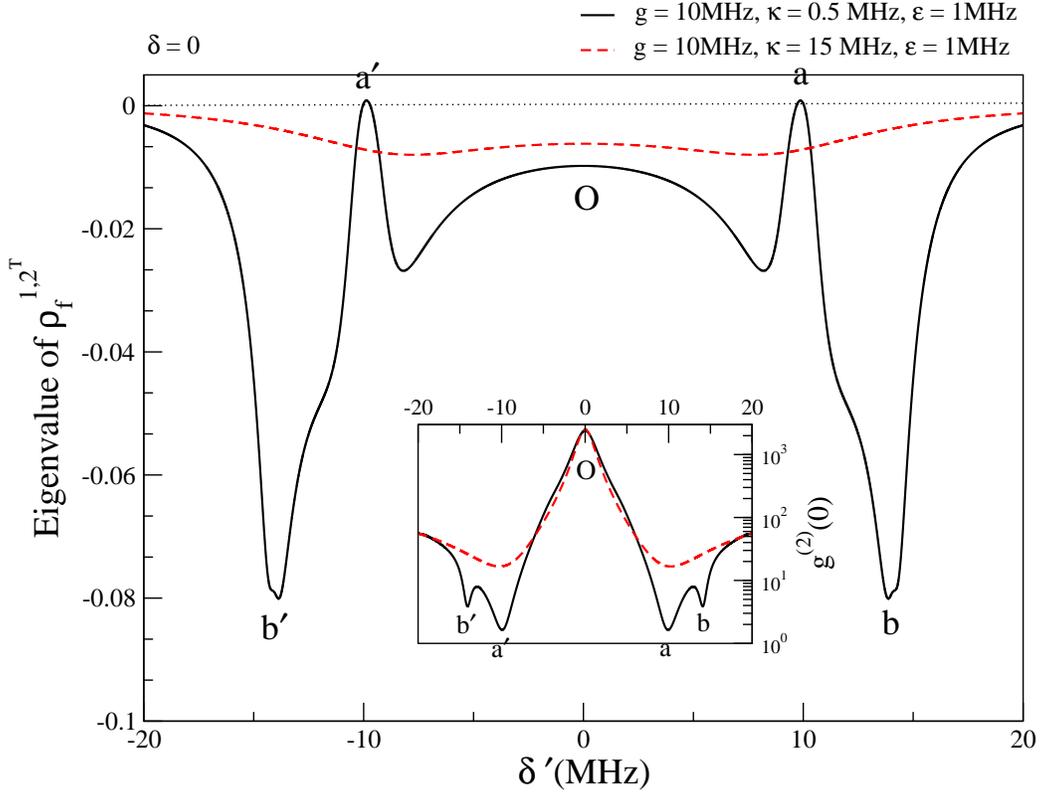}
\caption{(Color online) The lowest eigenvalue of the partial tranposed field density matrix $\rho ^{1,2^T}_f$ is plotted for strong (black solid) and weak (red dashed) coupling regimes, with detuning ($\delta'$) of one drive laser, keeping the other drive laser resonant to transition ($\delta = 0$). The system parameters are mentioned in the header. In the inset we have plotted the $g^{(2)}(0)$ as a function of $\delta'$ as in Fig.\ref{g20_detuning} to make a direct correspondence with the negativity of the eigenvalue. The black dotted line indicate zero value in y-axis.}
\label{par_trace}
 \end{figure}

The above observation prompts us to look into the entanglement aspects of the field modes. The necessary and sufficient condition for entanglement in any bipartite system is the negativity of at least one of the eigenvalues of the partial transpose \cite{peres:1996} of the density matrix of a bipartite system. The density matrix of our system is $\rho_{sf}$, where $sf$ stands for system-field. By doing partial trace over system basis states, we find reduced field density matrix $\rho^{1,2}_f$.
\be
\rho^{1,2}_f = Tr_s [\rho_{sf}]
\ee
where $f$ stands for the field and 1,2 stand for two field modes. Under partial transpose of Peres and Horodecki over subsystem 2 (or 1), the density matrix can be formally be written as $\rho^{1,2^T}_f$ or ($\rho^{1^T,2}_f$). Partial transpose over field mode 2 implies that the basis operators of the bipartite density matrix change as $\mid mn\rangle \langle \mu \nu \mid \rightarrow \mid m \nu \rangle \langle \mu n\mid$. For our system the $\rho^{1,2^T}_f$ matrix is of dimension 4$\times$4 and we calculate its eigenvalues numerically for different range of parameters. We indeed find that for a range of parameters $\rho^{1,2^T}_f$ has negative determinant and hence there exists at least one negative eigenvalue. We find the lowest eigenvalue to be negative for most of the parameter regime and plot it as a function of detuning of one of the drive lasers keeping the other drive laser at resonance to respective transition. The results are shown in Fig.\ref{par_trace}.

In Fig.\ref{par_trace} we can observe that for whole range of parameters except at positions ``a'' and ``a$'$'' where $\delta' = \pm {\rm g}$, the eigenvalue is negative. Negative eigenvalue implies that the field modes are indeed entangled. The negativity of the eigenvalue is prominent at $\delta' = \pm \sqrt{2}{\rm g}$, where photon induced tunneling take place, as describerd earlier. In the inset we can see that at positions ``b'' and ``b$'$'', $g^{(2)}(0)\sim 4$, i.e. the  photons are bunched and non thermal. In Fig.\ref{2pustrong}(iii) we can see that $C^{(2)} (0)$ has two peaks at $\delta' = \pm \sqrt{2}{\rm g}$, hence the photon pairs absorbed or emitted at this condition are entangled.  Since the two modes in our model can be nondegenerate in general, the two photons may be entangled in both frequency \cite{frequency} and polarization domains. This means that it may be possible to generate hyperentangled two-photon states \cite{hyperentanglement} by our model. Hyperentanglement is required in dense 
coding \cite{quantum_dense_coding} in quantum communication. However, this observation is evident only at strong coupling regime. On the other hand, at position ``O'' the $g^{(2)}(0)\sim 2000$, such high values arise from the low normalization denominator. Nevertheless, the eigenvalue is also negative at this position and importantly this negativity sustains in weak coupling regime as well. In the Fig.\ref{par_trace} at position ``a'' and ``a$'$'' when $\delta' = \pm {\rm g}$, $g^{(2)}(0)\sim 1.5$, i.e correlation is for all classical lights. The lowest eigenvalue of the partial transpose of the reduced density matrix also become positive here, implying the field modes are non entangled.

\begin{figure}
\center
 \includegraphics[width=\linewidth]{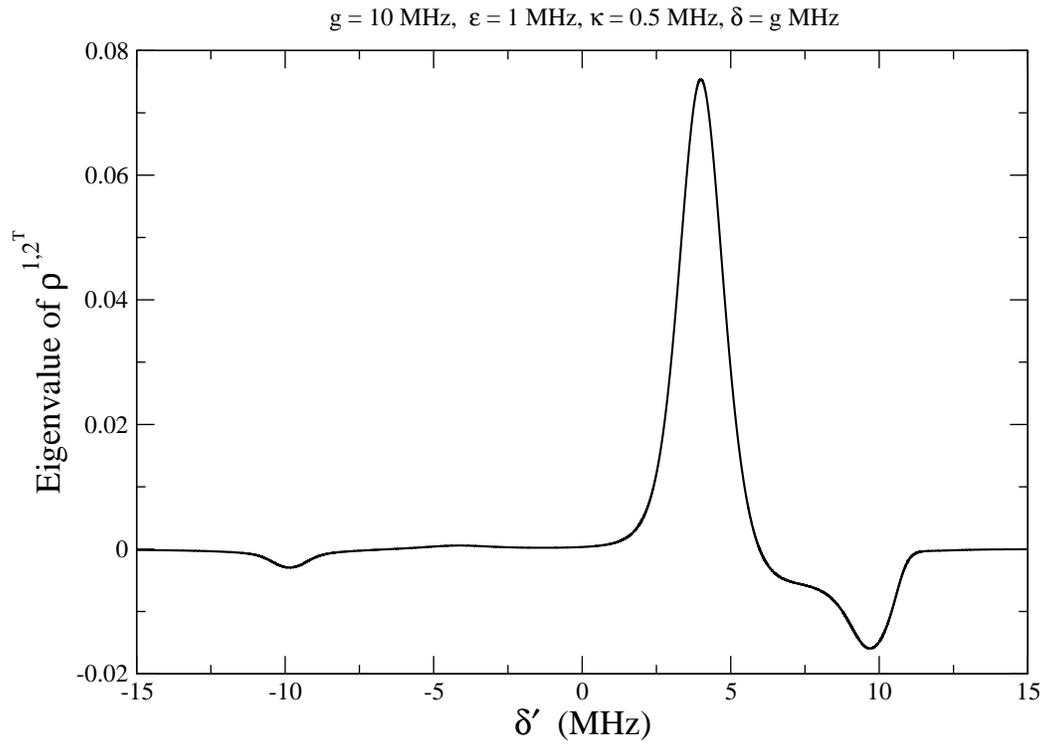}
\caption{The lowest eigenvalue of the partial tranposed field density matrix $\rho ^{1,2^T}_f$ is plotted for strong coupling regimes, with detuning ($\delta'$) of one drive laser, keeping $\delta = {\rm g}$. The system parameters are mentioned in the header. }
\label{par_trace_blockade}
\end{figure}

At $\delta' = \pm {\rm g}$ and $\delta ={\rm g}$ in strong coupling regime (see Fig. \ref{blockade}) the system shows antibunching of photons ,i.e. $g{(0)}<1$ and $C^{(2)}(0)<0$. At this very condition, the entanglement measure as described earlier shows the presence of negative eigenvalue (See Fig.
\ref{par_trace_blockade}), hence the field modes are entangled. But when $\delta' = (\sqrt{2} -1 ){\rm g}$ and $\delta ={\rm g}$, where the photon induced tunneling takes place, $g^{(2)}(0) \sim 1.5$ and $C^{(2)}(0)>0$ (Fig.\ref{blockade}), we find that all of the eigenvalues become positive and the lowest eigenvalue is plotted against $\delta'$ in Fig.
\ref{par_trace_blockade}, showing the positivity of the eigenvalue at $\delta' = (\sqrt{2} -1 ){\rm g}$. Hence the photons involved in the process for the stated parameters are non entangled because the correlation value $1<g^{(2)}(0)<2$.

\section{Conclusions}
In conclusion, we have studied the photon-photon correlations in detail for a $V$-type three level system in a two-mode cavity QED setup. We have shown that the anharmonicity of the dressed energy levels at low photon number sector induce effective photonic nonlinearity into the system. Our results show that the quantum fluctuations in the two modes that are coupled to a $V$-system become highly correlated $g^{(2)}(0)> 2$ under certain detuning of the drive fields. The negativity of the lowest eigenvalue of the partially transposed reduced field density matrix reveal that depending on detuning parameters when $g^{(2)}(0) > 2$, the field modes become entangled. We observe that, with red and blue detuned drive lasers at detuning ${\rm g}$ and $-{\rm g}$ in strong coupling regime, one can observe strong anticorrelation between the two photons ($g^{(2)}(0) \ll 1 $ or $C^{(2)}(0) \ll 0$), suggesting that the system may act as a generator of single photons. Hence the $V$-type system with long-lived or metastable excited states in CQED can behave either as correlated photon pair generator or a single photon source depending on suitable tuning of the drive lasers. The model described here may have potential applications in the fields of quantum state engineering, quantum information and communication. In addition this model may be extended to produce correlated photon laser \cite{zubairy:pra1987} by incoherently driving the $V$-system \cite{zubairy:pra2008}. For practical implementation of our model, cold fermionic $^{171}$Yb atoms may serve as a good candidate. Atomic $^{171}$Yb has ground state $^1S_0$ and metastable excited state $^3P_2$ and long-lived excited states $^3P_1$ which have lifetimes about $\sim$ 1  and $10^{-3}$ second, respectively.




\vspace{0.5cm} 

\noindent
{\bf Acknowledgment} \\
We thank Dr. Saikat Ghosh, IIT Kanpur,  for stimulating and helpful discussions.

\end{document}